\documentclass[12pt]{article}
\usepackage{amssymb}
\def\inh{\vskip 0.075truein \noindent\hangindent=12 pt \hangafter=1}

\usepackage{amsthm}\theoremstyle{remark}

\newcommand{\bte}{\begin{quote}\begin{theorem}}
\newcommand{\ete}[1]{\label{#1}\end{theorem}\end{quote}}
\newcommand{\bcom}{\begin{quote}\begin{comment}}
\newcommand{\ecom}[1]{\label{#1}\end{comment}\end{quote}}
\newcommand{\bex}{\begin{quote}\begin{example}}
\newcommand{\eex}[1]{\label{#1}\end{example}\end{quote}}
\newcommand{\bcon}{\begin{quote}\begin{conclusion}}
\newcommand{\econ}[1]{\label{#1}\end{conclusion}\end{quote}}
\newcommand{\bdefi}{\begin{quote}\begin{definition}}
\newcommand{\edefi}[1]{\label{#1}\end{definition}\end{quote}}

\newcommand{\blem}{\begin{quote}\begin{lemma}}
\newcommand{\elem}[1]{\label{#1}\end{lemma}\end{quote}}

\newcommand{\bpr}{\begin{quote}\begin{problem}}
\newcommand{\epr}[1]{\label{#1}\end{problem}\end{quote}}

\newcommand{\f}{\frac}
\newcommand{\p}{\partial}
\newcommand{\n}{\nonumber \\}
\newcommand{\inti}{\int_{-\infty}^\infty}
\newcommand{\beq}{\begin{eqnarray}}
\newcommand{\eeq}[1]{\label{#1}\end{eqnarray}}
\newcommand\eq[1]{(\ref{#1})}
\newcommand{\bfi}{\begin{figure}[24]}
\newcommand{\efi}[1]{\caption{\label{#1}}\end{figure}}
\newcommand\fig[1]{Fig.~\ref{#1}}
\newcommand{\res}{respectively}
\newcommand\gl{\left}
\newcommand\gr{\right}
\newcommand{\bfm}[1]{\mbox{\boldmath ${#1}$}}
\newcommand{\CE}{{\cal E}}

\newcommand{\CK}{{\cal K}}
\newcommand{\CL}{{\cal L}}

\newcommand{\CP}{{\cal P}}

\newcommand{\Gd}{\delta}
\newcommand{\Ge}{\varepsilon}

\newcommand{\Gf}{\phi}

\newcommand{\Gl}{\lambda}
\newcommand{\Gn}{\eta}
\newcommand{\Gm}{\mu}

\newcommand{\Gj}{\tau}

\newcommand{\Go}{\omega}

\newcommand{\GO}{\Omega}

\newcommand{\az}[1]{Sect.$\!$ \ref{#1}}
\newcommand\D{\,\mathrm{d}}
\newcommand\I{\mathrm{i}}
\newcommand\E{\mathrm{e}}
\newcommand{\bexe}{\begin{quote}\begin{exercise}\inh}
\newcommand{\eexe}[1]{\label{#1}\end{exercise}\end{quote}}

\newcommand{\la}{\langle}
\newcommand{\ra}{\rangle}
\usepackage{graphics,graphicx}
\usepackage{color}
\begin{document}

{\large
\title{On the energy partition in oscillations and waves}
}

\author{Leonid Slepyan}
\date{\small{{\em School of Mechanical Engineering, Tel Aviv University\\
P.O. Box 39040, Ramat Aviv 69978 Tel Aviv, Israel}
}}

\maketitle

\vspace{0mm}\noindent
{\bf Abstract}
A class of generally nonlinear dynamical systems is considered, for which the Lagrangian is represented as a sum of homogeneous functions of the displacements and their derivatives. It is shown that an energy partition as a single relation follows directly from the  Euler-Lagrange equation in its general form. It is defined solely by the homogeneity orders. If the potential energy is represented by a single homogeneous function, as well as the kinetic energy, the partition between these energies is defined uniquely. Finite discrete systems, finite continual bodies, homogeneous and periodic-structure waveguides are considered. The general results are illustrated by examples of various types of oscillations and waves, linear and nonlinear, homogeneous and forced, steady-state and transient, periodic, non-periodic and solitary, regular, parametric and resonant. The reduced energy partition relation for statics is also presented.

\section{Introduction}
For free linear oscillations and sinusoidal waves, it is long recognised that kinetic and potential energies averaged over the period are equal to each other. In some partial cases, this is observed directly, see, e.g, [1],[2]. This fact was also confirmed by Whitham [3] for a non-specified linear sinusoidal wave. This was found as a `side effect' from his version of the variation principle, which was introduced as a base for the theory of slow-varying sinusoidal waves. In the Whitham's considerations, the expression for a sinusoidal wave
\beq \Gf \sim \Re A\E^{\I(\bfm{k}\bfm{x} -\Go t)}\,,\eeq{i1}
where $A, \bfm{k}$ and $\Go$ are the slow varying parameters, is substituted in the Lagrangian neglecting derivatives of these parameters.
Then the Lagrangian is averaged over the period. Thereafter, the equipartition directly follows as a result of the variation of the wave amplitude.

 Along with this, the energy partition in different areas of nonlinear dynamics and in different sense is of interest, see, e.g., [1], [4], [5]. Besides, it seems that even the linear case not fully explored. For example, the cases of systems with time-dependent parameters, forced motions, solitary waves apparently are not discussed.

We consider this topic in terms of classical mechanics and obtain the partition relations for generally nonlinear oscillations and waves using a similar but not the same way. We consider a class of oscillations and waves, where the Lagrangian is represented by a sum of homogeneous functions of the displacements and their derivatives, possibly not only the first order. The Lagrangian can also depend on time and spatial coordinates explicitly. Starting from the Euler-Lagrange equation of a general view and taking into account Euler's theorem on homogeneous functions we obtain both the energy partition relation and the conditions defining the regions of averaging (it can be the oscillation period, if exists, and not only it). The partition relation is fully defined by the homogeneity orders regardless of other parameters of the system and the dynamic process. In the case, where the potential energy is represented by a single homogeneous function, as well as the kinetic energy, the partition between these energies is defined uniquely.

First, we consider general relations for finite discrete systems. The relations obtained for the latter are applicable, with minor additions, to a finite continuous-material body and to an infinite continuous or discrete waveguide as discussed below. Then some examples of free and forced oscillations are presented, which evidence that the linearity is neither necessary nor sufficient condition for the equipartition. Next, some examples of linear and nonlinear oscillations are presented, where the energy partition and the regions of averaging are clearly seen. Homogeneous and forced, steady-state and transient, periodic and non-periodic, regular, parametric and resonant oscillations are examined.

Further, the general relations are presented to be applicable for a finite continuous body and for homogeneous, periodic-structure and discrete waveguides. Some examples of linear and nonlinear, periodic, steady-state, transient and solitary waves are presented, where the energy partition is also demonstrated.

\section{The main relations}\label{mr}
Consider  the Euler-Lagrange equations
\beq \f{\p L}{\p u_i}-\f{\D}{\D t} \f{\p L}{\p \dot{u}_i} =0\,,~~~L=L(\bfm{u},\dot{\bfm{u}},t)\,, ~~~i=1, 2, ...\,,\eeq{001}
where $\bfm{u}=(u_1, u_2, ...)$. Multiplying the equations by $u_i$ and integrating over an arbitrary segment, $t_1\le t\le t_2$, we obtain (with summation on repeated indices)
\beq \int_{t_1}^{t_2} \gl( \f{\p L}{\p u_i}u_i +\f{\p L}{\p \dot{u}_i}\dot{u}_i\gr)\D t +B_1-B_2=0\,,\eeq{002}
where
\beq B_{1,2} = \f{\p L}{\p \dot{u}_i}{u}_i ~~~(t=t_{1,2})\,.\eeq{003}
Thus
\beq \int_{t_1}^{t_2} \gl( \f{\p L}{\p u_i}u_i +\f{\p L}{\p \dot{u}_i}\dot{u}_i\gr)\D t =0\eeq{004}
if
\beq B_1 = B_2\,.\eeq{005}

Regarding this formulation, we note that the relation \eq{004} follows from Hamilton's principle of least action with the variation $\Gd \bfm{u}$ replaced by $\bfm{u}$. This, however, entails a change in the additional conditions. Namely, in the variational formulation, the integration limits, $t_{1,2}$, are arbitrary under the condition $\Gd\bfm{u}=0$ at $t=t_{1,2}$, while in the modified formulation they are not arbitrary but must meet the condition \eq{005}.

We now suppose that the Lagrangian is a sum of homogeneous functions of $\bfm{u}, \dot{\bfm{u}}$
\beq L(\Gl\bfm{u},\Gl\dot{\bfm{u}},t) =  \Gl^{\nu_n} L_n(\bfm{u},\dot{\bfm{u}},t)\,,\eeq{006}
where $\nu_n$ is the homogeneity order. In this expression, $\nu_n$ can be not only integer, and it is assumed that only a positive value can have a fractional exponent. For example, in the latter case, $|u_i|^\nu$ can be present in the Lagrangian but not $u_i^\nu$. Accordingly, in the homogeneity definition, we take $\Gl>0$. It is also assumed that $|u_i|^\nu\ge 0$. Note that the homogeneity condition excludes possible non-uniqueness in the definition of the energies.

With refer to Euler's theorem on homogeneous functions
\beq \f{\p f(\bfm{x})}{\p x_i}x_i = \nu f(\bfm{x})~~~\mbox{if}~~~f(\Gl\bfm{x})=\Gl^\nu f(\bfm{x})\eeq{EThf1}
and the condition \eq{005} we can rewrite the relation \eq{004} in the form
\beq \int_{t_1}^{t_2} \nu_n L_n\D t =0\,.\eeq{007}

The latter relation can serve for the determination of the energy partition.
For example, let $L$ be a difference between the kinetic and potential energies,
$L=\CK-\CP$, where the, $\CK$ and $\CP$, are homogeneous functions of the orders $\Gm$ and $\nu$, \res. In this case, the relation between the averaged energies
\beq \langle \CK\rangle = \f{1}{\Gj}\int_{t_1}^{t_1+\Gj} \CK\D t\,,~~~  \langle \CP\rangle = \f{1}{\Gj}\int_{t_1}^{t_1+\Gj} \CP\D t~~~(\Gj=t_2-t_1)\eeq{008}
is
\beq
\f{\langle \CK\rangle }{\langle \CP\rangle }= \f{\nu}{\Gm} ~~~\Longrightarrow ~~~\langle \CK\rangle =\f{\nu \CE}{\Gm+\nu}\,,~~
\langle \CP\rangle =\f{\Gm \CE}{\Gm+\nu}\,,\eeq{009}
where $\CE=\CK+\CP$ is the total energy.

In particular, in the linear case, where $\nu=\Gm=2$, the averaged energies are equal.
Cases $\nu> \Gm$ and $\nu <\Gm $ correspond to hardening and softening nonlinearity, \res. The potential energy vanishes as $\nu/\Gm \to \infty$. In the limit, this case corresponds to the periodic perfect collisions of rigid particles.

Note that {\bf in statics}, $\CK=0$, the boundary terms, $B_{1,2}$, vanish independently of $t_{1,2}$. It follows from \eq{007} that
\beq \nu_n \CP_n =0\,.\eeq{sccp0}
Thus, the energy is at zero if it is a homogeneous function; otherwise, if it is a sum of such functions this relation defines the energy partition (in this connection, see an example in \az{P4}). 

It is remarkable that the energy ratio in \eq{009} is equal to the homogeneity order ratio regardless of the other parameters of the system and the dynamic process. However, in the case where the potential energy is represented by a sum of two or more homogeneous functions of different orders, the relation \eq{007} is only one for three or more terms. In this case, it is not sufficient for the determination of ratios between all terms but still allows one to make some conclusions concerning the energy partition.

The condition in \eq{005} is satisfied in periodic oscillations
\beq \bfm{u}(t+\Gj )=\bfm{u}( t)\eeq{010}
and also in the case where $\bfm{u}=0$ at $t=t_{1,2}$. In addition, if the kinetic energy being a homogeneous function of order $\mu$ with respect to $u_i$, $\dot{u_i}$ is such a function of order 2 with respect to $\dot{u}_i$
\beq \CK(\Gl\bfm{u}, \Gl \bfm{\dot{u}})=\Gl^\mu\CK(\bfm{u}, \bfm{\dot{u}})\,,~~~\CK(\bfm{u}, \Gl \bfm{\dot{u}})=\Gl^2\CK(\bfm{u}, \bfm{\dot{u}})\,,\eeq{011}
then
\beq B_{1,2} = B^0_{1,2}u_i\dot{u}_i\eeq{012}
and $B_1 (B_2)=0$ if $\dot{\bfm{u}}=0$ at $t=t_1 (t_2)$. Therefore the condition \eq{005} is also satisfied in the case of non-periodic oscillations, where $\bfm{u}$ or/and $\bfm{\dot{u}}$ vanish at points. This can be seen in the below examples of oscillations and periodic and solitary waves.
In the case of regular or irregular oscillations, does not matter, where the condition \eq{005} is satisfied for each of the neighboring points, $t_1< t_2< ... <t_m < ...$, the energy partition averaged over any interval between two of such points is the same. So the energy partition appears fixed for a large range of time.

In this section, we have considered systems of a finite number of material points. The main relations presented here are applicable, with minor additions, to a finite continuous-material body and to a infinite continuous or discrete waveguide as discussed in \az{fcb} and \az{wave}.

While the energy partition is defined solely by the homogeneity orders, below we calculate the energy distribution function, $\CL(t)$, as a function of time
\beq \CL(t)=\int_0^t \nu_n L_n(t)\D t\eeq{Lot}
assuming that $B_1=0$ at $t=t_1=0$. This function must vanish at $t=t_2$ \eq{007} if the condition \eq{005} is satisfied at that point.
 This representation allows us to demonstrate how the partition varies during the $(t_1, t_2)$-interval and how $\CL(t)$ follow the statement \eq{007}  (also see the expression in \eq{012}). As a rule, dimensionless quantities are used since this simplification does not affect the partition.

\section{Discrete systems}\label{KaPEs}

\subsection{The linearity and equipartition}
\begin{figure}[h!]

\vspace{2mm}
\hspace{30mm}\includegraphics*[width=.6\textwidth]{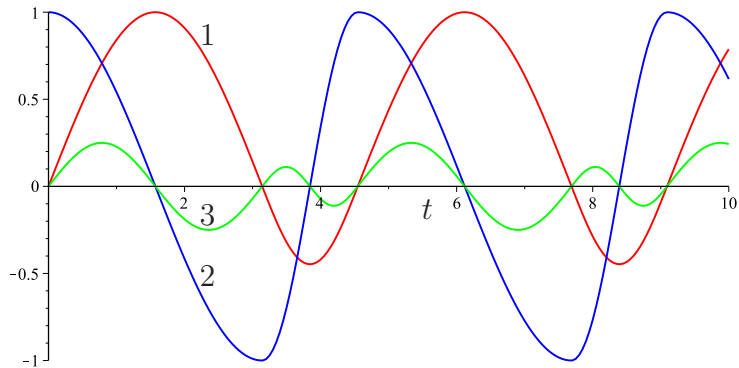}

\begin{picture}(0,0)(0,-100)
\put(161,41){1}
\put(161,-50){2}
\put(161,-27){3}
\put(245,-25){$t$}
\end{picture}

\vspace{-4mm}
\caption[]{Peace-linear oscillator. Displacements $u(t)$ (1), speed $\dot{u}(t)$ (2) and the energy distribution function $\CL(t)/2$ (3) for  $\CP_+=1, \CP_- =5$. It is seen that $\CL=0$ at points where $u=0$ and where $\dot{u}=0$.}
\label{osc5}
\end{figure}

Three examples are presented below which evidence that the linearity is neither necessary nor sufficient condition for the equipartition. First, consider a piecewise linear equation
\beq \ddot{u}(t) +[\CP_+H(u)+\CP_-H(-u)]u(t)=0\,,\eeq{lanoto1}
where $H$ is the Heaviside step function and $\CP_\pm$ are different positive constants. The corresponding Lagrangian, kinetic and potential energies and homogeneity orders are, \res
\beq L=\CK-\CP\,,~~~\CK(t) =\f{m\dot{u}^2(t)}{2}\,,~~~\CP(t)=[\CP_+H(u)+\CP_-H(-u)]\f{u^2(t)}{2}\,,~~~\Gm=\nu=2\,.\eeq{eqp1}
Equation \eq{lanoto1} defines periodic oscillations. The displacement, $u(t)$,  speed $\dot{u}(t)$ and the energy distribution function $\CL(t)$ \eq{Lot}
are presented in \fig{osc5} for $\CP_+=1, \CP_-=5$ and initial conditions $u(0)=0, \dot{u}(0)=1$.

It can be seen that the energies averaged over the period, $\la \CK\ra$ and $ \la \CP\ra$ (in this vase, period $\Gj=(1+1/\sqrt{5})\pi\approx 4.5465556$), are equal:
\beq \CL(\Gj)=\int_0^\Gj 2L(t)\D t=\int_t^{t+\Gj} 2L(t)\D t=2\Gj[\la \CK\ra - \la \CP\ra] =0\,.\eeq{lanoto2}

\begin{figure}[h!]

\vspace{2mm}
\hspace{30mm}\includegraphics*[width=.4\textwidth]{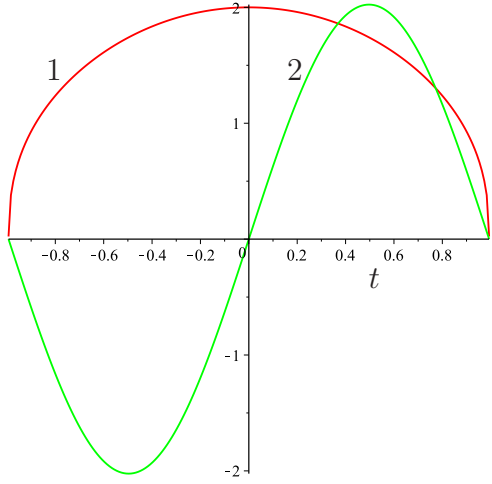}

\begin{picture}(0,0)(0,-100)
\put(194,69){2}
\put(103,69){1}
\put(225,-10){$t$}
\end{picture}

\vspace{-4mm}
\caption[]{The bubble radius, $r(t)$ (1) and the energy distribution function $\CL(t)/2$ (2)}
\label{bub1}
\end{figure}

Next, we consider a nonlinear equation related to centrosymmetric oscillations of a bubble in unbounded, perfect, incompressible liquid, where the added mass is proportional to the bubble radius cubed. Let the (dimensionless) energies be
\beq \CK=\f{1}{2}r^3(t)\dot{r}^2(t)\,,~~~\CP =\f{1}{5} r^5(t)\,,~~~\Gm=\nu=5~~~(\mbox{the radius}\, \, r(t)\ge 0)\,.\eeq{bub1}
The corresponding dynamic equation is
\beq r(t)\ddot{r}(t) +\f{3}{2}\dot{r}^2(t)+r(t)^2=0\,.\eeq{bub2}
The radius, $r(t)$ and the energy distribution function $\CL(t)$ under conditions $r(0)=2, \dot{r}(0)=0$ are plotted in \fig{bub1}

It is seen that the equipartition does take place in this strongly nonlinear situation as well.

\begin{figure}[h!]

\vspace{-2mm}
\hspace{30mm}\includegraphics*[width=.6\textwidth]{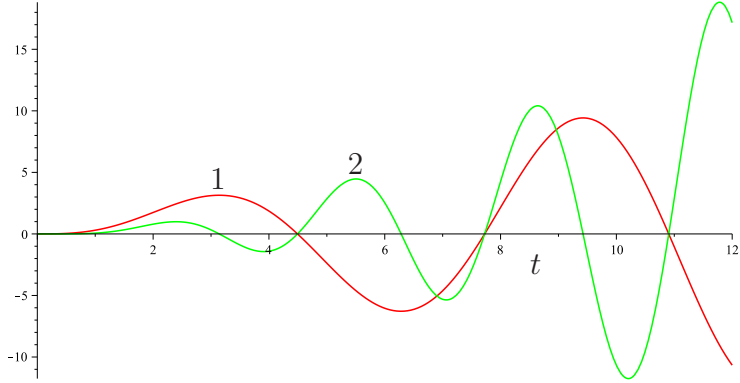}

\begin{picture}(0,0)(0,-100)
\put(217,-5){2}
\put(165,-11){1}
\put(286,-43){$t$}
\end{picture}

\vspace{-4mm}
\caption[]{Inhomogeneous problem: The ordinary resonance. The displacement $u(t)$ (1) and the energy distribution function $\CL(t)$ (2). It is seen that zeros of the latter correspond to zeros of $u(t)\dot{u}(t)$.}
\label{res}
\end{figure}

Next consider forced resonant oscillations as an inhomogeneous linear problem
\beq \ddot{u}+u=P\sin t\,,~~~
\CK = \f{\dot{u}^2}{2}~~(\Gm=2)\,,\n\CP=\CP_1 +\CP_2\,,~~~\CP_1 = \f{u^2}{2}~~(\nu_1=2)\,,~~~\CP_2=- Pu\sin t~~(\nu_2=1)\,.\eeq{fro1}
It follows that under zero initial conditions
\beq u=\f{P}{2}(\sin t -t\cos t)\,. \eeq{fro2}
The displacements and the energy distribution function $\CL(t)$ are plotted in \fig{res} for $P=1$.

Thus, there is no equipartition between the kinetic and potential energies in this inhomogeneous linear problem:
\beq \la \CK\ra = \la \CP_1\ra +\f{1}{2}\la\CP_2\ra < \la \CP\ra =\la \CP_1\ra +\la \CP_2\ra\,.\eeq{ldnre1}

\subsection{Simplest example of unequal partition}
Consider a ball of mass $M$ thrown upwards with speed $\dot{u}(0)=v_0$. The energies are
\beq \CK = \f{M\dot{u}^2}{2}\,,~~~\CP =Mgu~~~(\Gm=2\,,~\nu=1)\,,\eeq{btu1}
where $g$ is the acceleration of gravity. Here $u=0$ at $t=t_1=0$ and $\dot{u}=0$ at $t=t_2=v_0/g$. It follows from \eq{007} that the energies averaged over the $(t_1,t_2)$-segment must satisfy the relation
\beq \la \CP\ra = 2\la \CK\ra\,,\eeq{btu2}
which can also be obtained by direct calculations. Indeed,
\beq \dot{u}=v_0 - gt\,,~~u=v_0t-\f{gt^2}{2}\,,\n \la \CK\ra=\f{1}{2t_2}\int_0^{t_2}M(v_0 - gt)^2\D t =\f{Mv_2^2}{6}\,,\n
\la \CP\ra =\f{1}{t_2}\int_0^{t_2}Mg\gl(v_0t - \f{gt^2}{2}\gr)\D t =\f{M v_0^2}{3}=2\la \CK\ra\,.\eeq{btu3}

\subsection{Linear and nonlinear oscillators}
\begin{figure}[h!]

%\vspace{-60mm}
\centering
\vspace*{10mm} \rotatebox{0}
{\resizebox{!}{5.cm}{\includegraphics[scale=0.3]{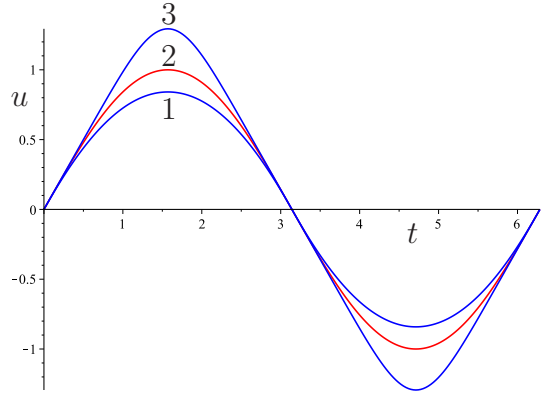}}}

\begin{picture}(0,0)(0,-100)
\put(-43,20){1}
\put(-43,40){2}
\put(-43,56){3}
\put(50,-27){$t$}
\put(-100,25){$u$}
\end{picture}

\vspace{-2mm}
\caption{The oscillators: $\nu = 6/5, \CP_0= 0.6152$ (1), $\nu= 2, \CP_0=1/2$ (2) and $\nu=6, \CP_0=0.1067$ (3). Displacements under initial conditions $u=0, \dot{u}=1$ for each of the oscillators.}
    \label{osc1}
\end{figure}

\begin{figure}[h!]

\vspace{-2mm}
\hspace{50mm}\includegraphics*[width=.4\textwidth]{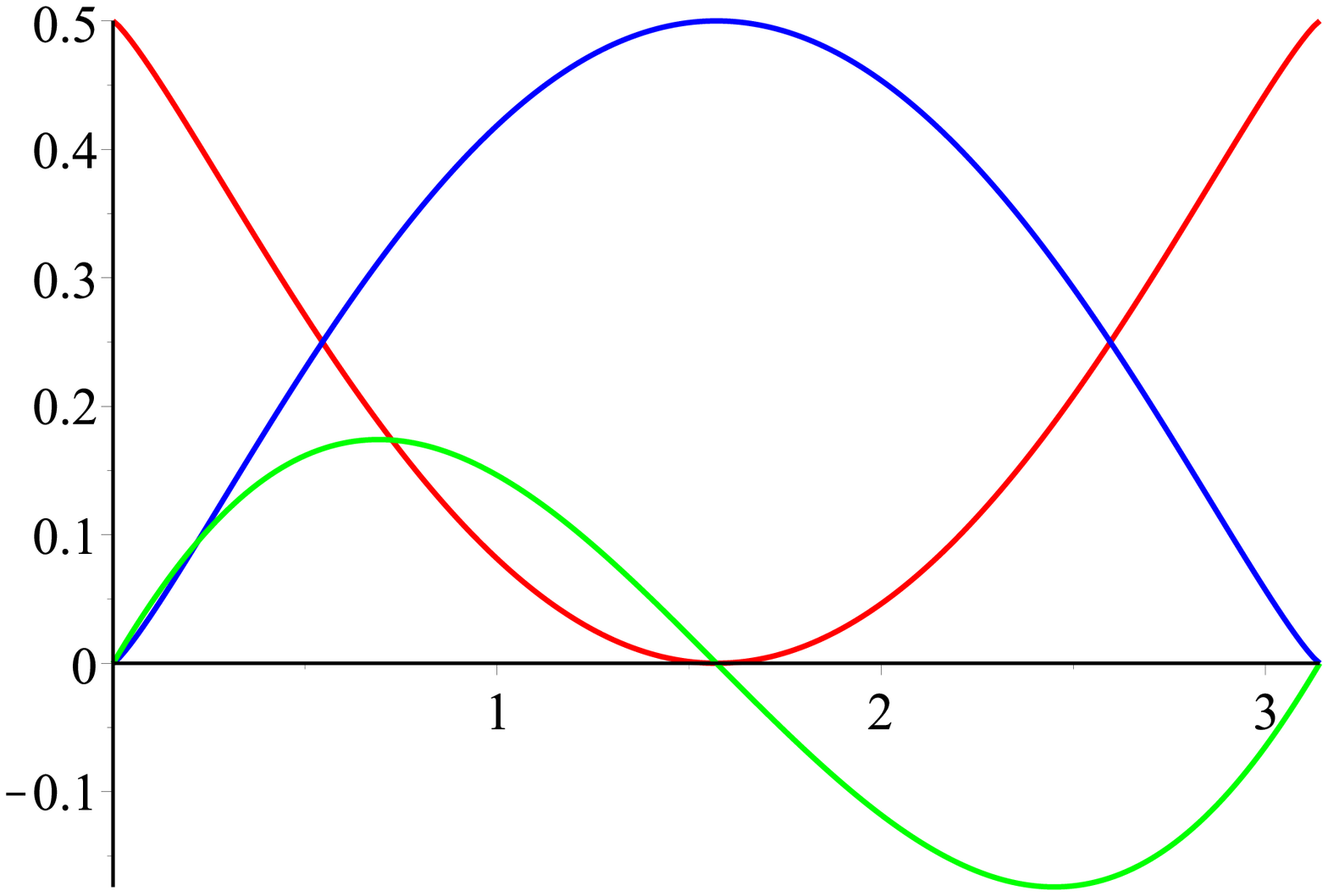}
\begin{picture}(0,0)(0,-100)
\put(-53,-39){1}
\put(-53,3){2}
\put(-53,-93){3}
\put(-33,-80){$t$}
\end{picture}

\vspace{0mm}
\caption[]{Nonlinear oscillator, $\nu=6/5, \CP_0=0.6152$. Kinetic energy (1), potential energy (2) and the energy distribution function $\CL(t)/2$ (3).}
\label{osc2}
\end{figure}

\begin{figure}[h!]

\vspace{5mm}
\hspace{50mm}\includegraphics*[width=.4\textwidth]{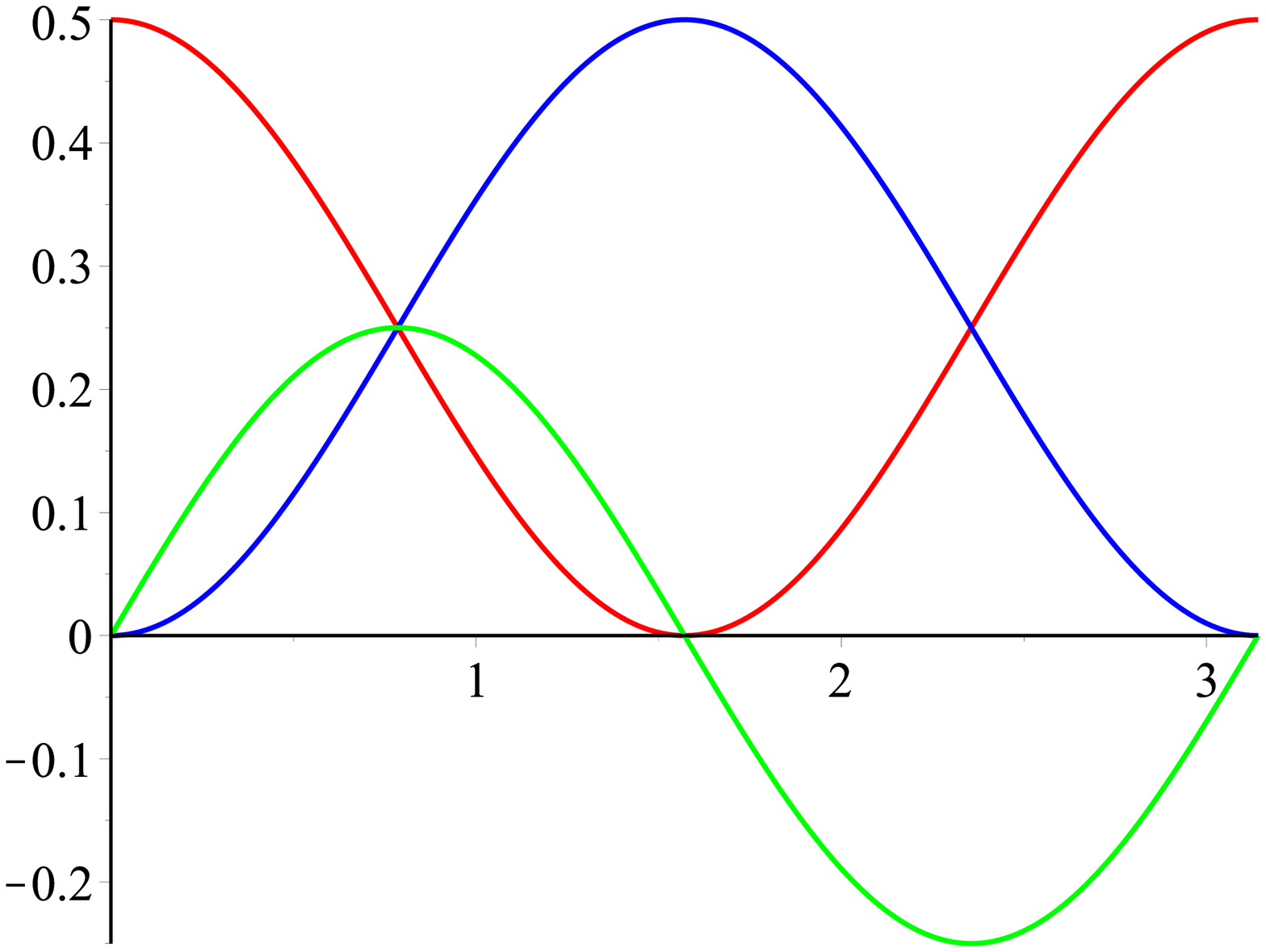}

\begin{picture}(0,0)(0,-100)
\put(268,-14){1}
\put(268,36){2}
\put(268,-70){3}
\put(300,-52){$t$}
\end{picture}

\vspace{-2mm}
\caption[]{Linear oscillator, $\nu=2, \CP_0=1/2$. Kinetic energy (1), potential energy (2) and the energy distribution function $\CL(t)/2$ (3).}
\label{osc3}
\end{figure}

\begin{figure}[h!]

\vspace{12mm}
\hspace{50mm}\includegraphics*[width=.4\textwidth]{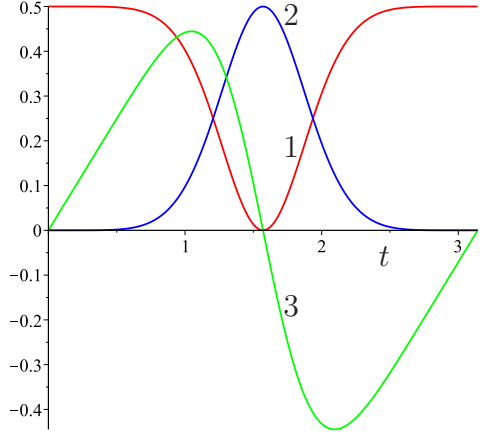}

\begin{picture}(0,0)(0,-100)
\put(254,21){1}
\put(254,71){2}
\put(254,-39){3}
\put(290,-20){$t$}
\end{picture}

\vspace{-2mm}
\caption[]{Nonlinear oscillator, $\nu=6, \CP_0=0.1067$. Kinetic energy (1), potential energy (2) and the energy distribution function $\CL(t)/2$ (3).}
\label{osc4}
\end{figure}

Consider oscillators, which kinetic and potential energies are
\beq \CK =m \f{\dot{u}^2}{2}\,,~~~\CP = \CP_\pm\gl(\f{|u|}{u_0}\gr)^{\nu}\,,\eeq{kapefo1}
where the coefficients $\CP_+$ and $\CP_-$ correspond to positive and negative values of $u$, \res, and can be different, $(m, \CP_\pm, u_0, \nu)>0$.
For these systems the ratio of the kinetic-to-potential averaged energies is equal to $\nu/2$. Plots of the displacements and energies for $\nu = 6/5, 2$ and $6$ are shown in \fig{osc1}$-$\fig{osc4}. To have the same period, $2\pi$, for all these cases we take $\CP_+=\CP_- = 0.6152, 1/2, 0.1067$ for $\nu=6/5, 2$ and $6$, \res, and $m= u_0 =1$, $\CE=1/2$ for each of them. It is seen that zeros of $\CL(t)$ correspond to the condition \eq{005} with \eq{012}. Note that an analytical solution exists corresponding to the above equation \eq{kapefo1}. However here we are not interested in it since the energy partition is defined solely by the homogeneity orders.

\subsection{Two-degree-of-freedom nonlinear system}
\begin{figure}[h!]

%\vspace{-60mm}
\centering
\vspace*{0mm} \rotatebox{0}
{\resizebox{!}{3.cm}{\includegraphics[scale=0.3]{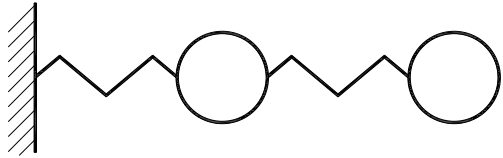}}}
\caption{Nonlinear mass-spring system of two degree of freedom.}
    \label{td-osc}
\end{figure}

Consider a mass-spring chain, \fig{td-osc}, which energies are
\beq \CK = \f{m_1\dot{u}_1^2}{2} + \f{m_2\dot{u}_2^2}{2}\,,~~~\CP = \CP_{1\pm}\gl(\f{|u_1|}{u_0}\gr)^\nu + \CP_{2\pm}\gl(\f{|u_1-u_2|}{u_0}\gr)^\nu\,.\eeq{tdf1}

\begin{figure}[h!]

%\vspace{-60mm}
\centering
\vspace*{10mm} \rotatebox{0}
{\resizebox{!}{5.cm}{\includegraphics[scale=0.4]{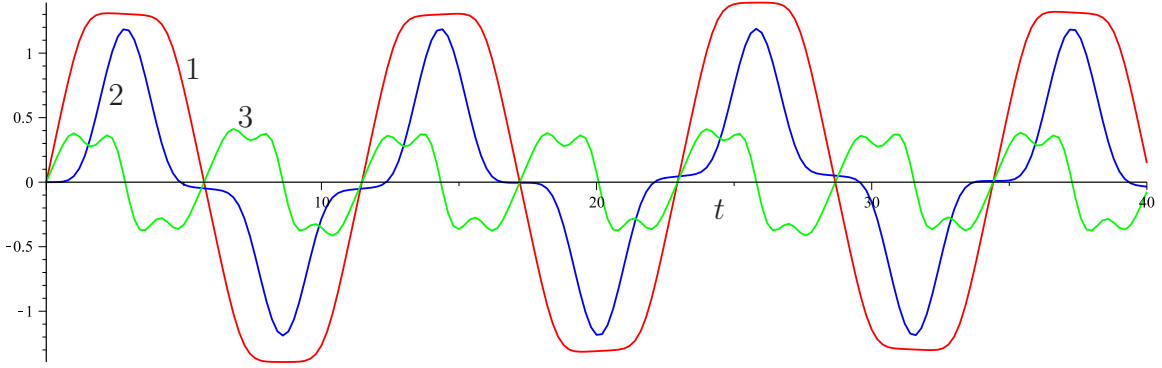}}}

\begin{picture}(0,0)(0,-100)
\put(-149,25){1}
\put(-178,15){2}
\put(-129,7){3}
\put(51,-28){$t$}
\end{picture}
\caption{Nonlinear mass-spring system of two degree of freedom. Displacements $u_1(t)$ (1), $u_2(t)$ (2) and $\CL(t)/2$ (3); $\nu=4$. Zeros of $\CL(t)$ appearing in accordance with condition \eq{005} with \eq{012} during the visible period can be observed.}
    \label{td-osc1}
\end{figure}

Plots of $u_1(t), u_2(t)$ and $\CL(t)$ for $m_1=m_2=u_0=1, \CP_1=\CP_2=1/4$ and $\nu=4$ based on the corresponding dynamic equations
\beq \f{\D^2 u_1}{\D t^2}= -(u_1-u_2)^3\,,~~~\f{\D^2 u_2}{\D t^2}= (u_1-u_2)^3-u_2^3\eeq{tdof3}
are shown in \fig{td-osc1}. An agreement with the relations \eq{007} and \eq{005} with \eq{012} is demonstrated.

\subsection{Oscillators under variable stiffness}
\begin{figure}[h!]

\vspace{-2mm}
\hspace{30mm}\includegraphics*[width=.6\textwidth]{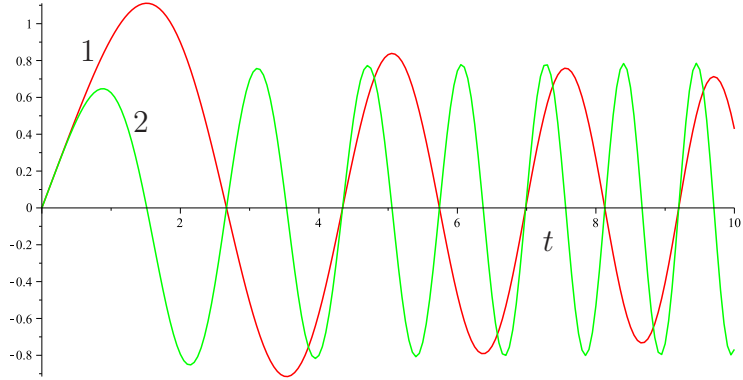}

\begin{picture}(0,0)(0,-100)
\put(135,10){2}
\put(115,37){1}
\put(290,-35){$t$}
\end{picture}

\vspace{-4mm}
\caption[]{Oscillator with linearly increasing stiffness. The displacement $u(t)$ (1) and the energy distribution function $\CL(t)$ (2). It is seen that zeros of the latter correspond to zeros of $u(t)\dot{u}(t)$.}
\label{tex}
\end{figure}

Consider oscillators which potential energy depends on $t$ explicitly. Let in the first one the energies be
\beq \CK=\f{\dot{u}^2}{2}\,,~~~\CK=\f{tu^2}{2}\,,~~~\nu=\Gm=2\,.\eeq{tex1}
It follows from \eq{004}  that, in spite of the aperiodicity of the oscillations, there is the energy equipartition.  It concerns the kinetic and potential energies averaged over any interval, $t_1,t_2$, where $B_1=B_2$ \eq{005}
The corresponding dynamic equation is
\beq \ddot{u}+t u =0\,.\eeq{tex2}
Its solution, $u(t)$, under conditions $u(0)=0, \dot{u}(0)=1$ and the energy distribution function $\CL(t)$ \eq{Lot} are plotted in \fig{tex}. It is seen that zeros of $\CL(t)$ coincide with zeros of $u(t)\dot{u}(t)$, as should be.

\begin{figure}[h!]

\vspace{-2mm}
\hspace{30mm}\includegraphics*[width=.6\textwidth]{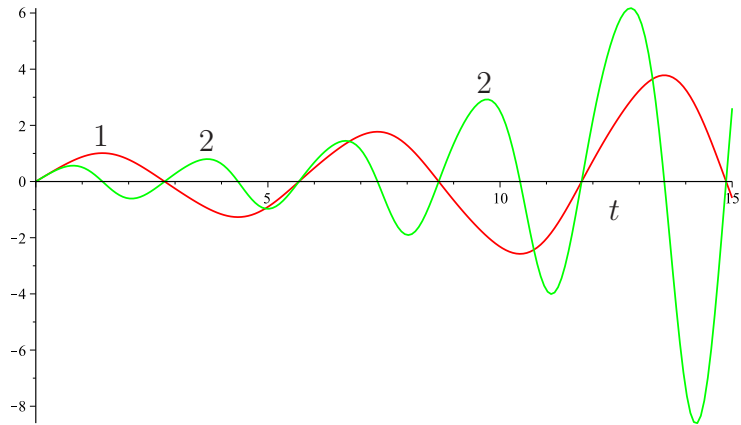}

\begin{picture}(0,0)(0,-100)
\put(160,21){2}
\put(120,23){1}
\put(265,43){$2$}
\put(315,-5){$t$}
\end{picture}

\vspace{-4mm}
\caption[]{Parametric resonance. The displacement $u(t)$ (1) and the energy distribution function $\CL(t)$ (2). It is seen that zeros of the latter correspond to zeros of $u(t)\dot{u}(t)$.}
\label{M1}
\end{figure}

Next, we consider parametric resonance based on  Mathieu's differential equation. The energies are
\beq \CK=\f{\dot{u}^2}{2}\,,~~~\CP=\f{(1+0.5\sin 2t)u^2}{2}\,,~~~\nu=\Gm=2\,.\eeq{M1}
The corresponding equation is
\beq \ddot{u} + (1+0.5\sin 2t)u =0\,.\eeq{M2}
Its transient solution, $u(t)$, under conditions $u(0)=0, \dot{u}(0)=1$ and the energy distribution function $\CL(t)$ \eq{Lot} are plotted in \fig{M1}. It is seen that zeros of $\CL(t)$ coincide with zeros of $u(t)\dot{u}(t)$, as should be.

\section{A finite continuous body}\label{fcb}
In this case, the Lagrangian depends, in addition, on  derivatives with respect to the spatial variables. The corresponding Euler-Lagrange equation multiplied by $\bfm{u}(\bfm{x},t)$ is now integrated by parts over both the time-segment ($t_1,t_2$) and the body material volume, $\GO$. As a result we obtain the relation \eq{007}, where $L$ is the Lagrangian incorporating the energy of the whole body including its boundaries, whereas the condition \eq{005} becomes
\beq \int_\GO \gl(B_2-B_1\gr)\D\GO =0\,.\eeq{coar1}
Note that the boundary terms, which can in integration by parts over $\GO$, reflect the energy associated with the body boundaries. So, in contrast to the difference in \eq{coar1}, they do not affect the basic relation \eq{007}.

\begin{figure}[h!]

\vspace{7mm}
\hspace{30mm}\includegraphics*[width=.4\textwidth]{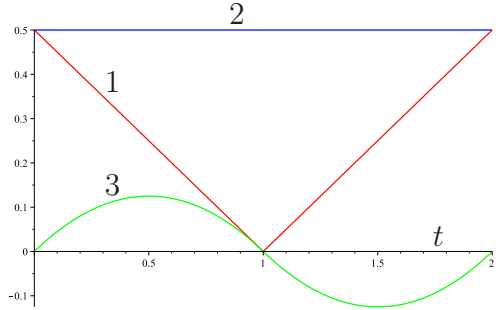}

\begin{picture}(0,0)(0,-100)
\put(171,24){2}
\put(124,-2){1}
\put(124,-41){3}
\put(248,-60){$t$}
\end{picture}

\vspace{-4mm}
\caption[]{Collision of a rod with a rigid obstacle. The kinetic energy (1), the total energy (2) and the energy distribution
 function $\CL(t)$ (3). Zeros of the latter are at $t=0$ ($u(x,0) =0$),  $t=1$ ($\dot{u}(x,1)=0$) and $t=2$ ($u(x,2) =0$). }
\label{bc}
\end{figure}

As an example consider the collision of a linearly elastic rod, $0<x<1$, with a rigid obstacle. For this linear problem the (dimensionless) energies are
\beq \CK =\f{1}{2}\int_0^1 \dot{u}^2 \D x\,,~~~\CP=\f{1}{2}\int_0^1 (u')^2 \D x\,,\eeq{coar3}
and the homogeneity orders $\mu=\nu=2$. The corresponding 1D wave equation is
\beq \ddot{u}(x,t)-u''(x,t)=0\,,~~~(0<x<1)\eeq{coar1cc}
with additional conditions
\beq u(x,0)=0\,, ~~\dot{u}(x,0)=1\,,~~ u(0,t)=0~~(0<t<2)\,,~~u'(1,t)=0\,.\eeq{coar2}
The collision period: $0<t<2$. Dependencies for the energies, $\CK$ and $\CP=\CE-\CK$ (the total energy $\CE=1/2$), and the energy distribution function $\CL(t)$ are plotted in \fig{bc}.

\begin{figure}[h!]

\vspace{-2mm}
\hspace{30mm}\includegraphics*[width=.6\textwidth]{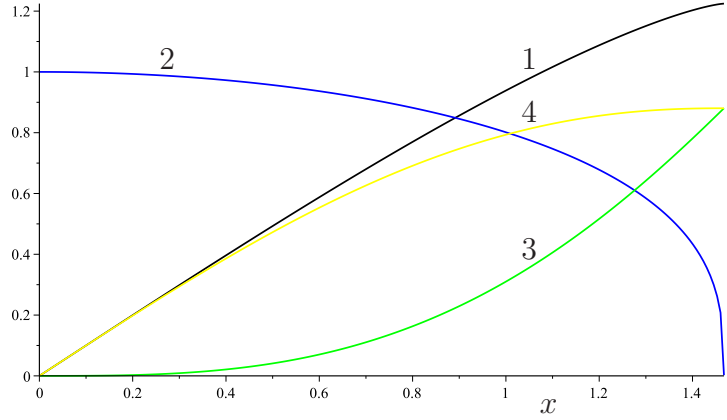}

\begin{picture}(0,0)(0,-100)
\put(151,43){2}
\put(288,43){1}
\put(288,-29){$3$}
\put(295,-87){$x$}
\put(288,23){4}
\end{picture}

\vspace{-4mm}
\caption[]{Oscillations of a nonlinearly elastic rod. Functions $X(x)$ (1),  $X'(x)$ (2), $\int_0^x X^2(x)\D x$ (3) and $\int_0^x (X')^4(x)\D x$ (4).}
\label{nr1}
\end{figure}

\begin{figure}[h!]

\vspace{-2mm}
\hspace{30mm}\includegraphics*[width=.6\textwidth]{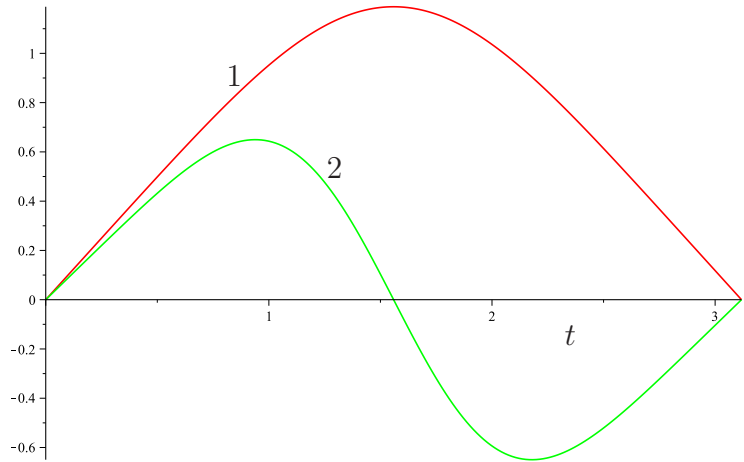}

\begin{picture}(0,0)(0,-100)
\put(170,59){1}
\put(208,24){$2$}
\put(298,-39){$t$}
\end{picture}

\vspace{-4mm}
\caption[]{Oscillations of a nonlinearly elastic rod. Functions $T(t)$ (1) and $\CL(t)$. (2).}
\label{nr2}
\end{figure}

\begin{figure}[h!]

\vspace{-2mm}
\hspace{30mm}\includegraphics*[width=.6\textwidth]{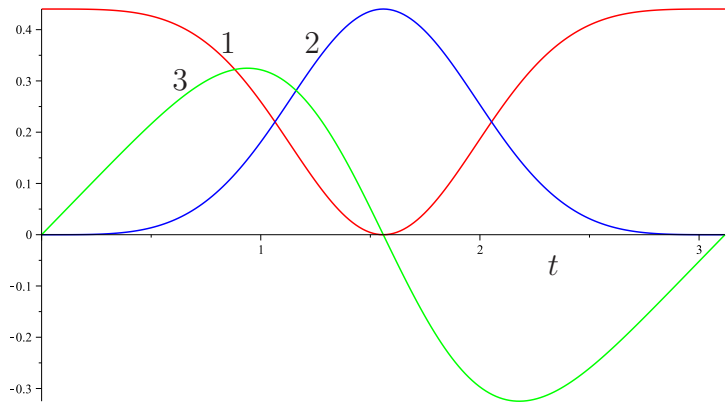}

\begin{picture}(0,0)(0,-100)
\put(206,49){2}
\put(174,49){1}
\put(156,35){$3$}
\put(298,-35){$t$}
\end{picture}

\vspace{-4mm}
\caption[]{Oscillations of a nonlinearly elastic rod. The kinetic energy (1), potential energy (2) and $\CL(t)$ (3).}
\label{nr3}
\end{figure}

Next, we consider oscillations of a nonlinearly elastic rod, $0<x<l$. Let the energies be
\beq \CK = \int_0^l \f{\dot{u}^2}{2}\D x\,,~~~\CP= \f{1}{2}\int_0^l \f{(u')^4}{4}\D x\,,~~~\Gm=2\,,~\nu=4\,.\eeq{coar4}
We solve the corresponding dynamic equation
\beq  \ddot{u}-3(u')^2u'' =0\eeq{coar4a}
by separation of variables, $u(x,t) = X(x)T(t)$, that results in equations
\beq (X')^2X'' +\f{\Gl}{3}X=0\,,~~~\ddot{T}+\Gl T^3=0\,,\eeq{coar5}
where $\Gl$ is an arbitrary constant. Note that these equations are solvable analytically. To simplify our considerations we take values of $\Gl$ and additional conditions and obtain the corresponding length of the rod and the oscillation amplitude. Namely, we take
\beq \Gl=1\,,~~ X(0)=X'(l)=T(0)=0\,,~~X'(0)=\dot{T}(0)=1\eeq{coar6}
and obtain the rod length and the oscillation period as
\beq l=1.4674161382\,,~~~\Gj =3.118169\,.\eeq{coar7}
The dependencies of the displacements and energies on the coordinate and time and the energy distribution function $\CL(t)$ are presented in \fig{nr1} $-$ \fig{nr3}.

\section{Waves}\label{wave}
In the case where the total energy of a wave is infinite, we have to choose a finite segment of  the waveguide $(x_1,x_2)$, similar to $(t_1,t_2)$ in time, to avoid boundary terms in integration by parts over this segment. Namely, in the conversions
\beq \int_{x_1}^{x_2}u_i\f{\D}{\D x}\f{\p L}{\p u'_i}\D x = -\int_{x_1}^{x_2} \f{\p L}{\p u'_i}u'_i\D x + \f{\p L}{\p u'_i} u_i|_{x_1}^{x_2}\,,\n
 \int_{x_1}^{x_2}u_i\f{\D^2}{\D x^2}\f{\p L}{\p u''_i}\D x = \int_{x_1}^{x_2} \f{\p L}{\p u''_i}u''_i\D x + \f{\D}{\D x}\f{\p L}{\p u''_i}u_i|_{x_1}^{x_2} -\f{\p L}{\p u''_i}u'_i |_{x_1}^{x_2}\eeq{ww1}
and so on, the segment should be chosen such that the boundary terms vanish. In this case, the energy partition relation \eq{007} remains valid with respect to the energies averaged over the space-time region ($x_1< x< x_2, t_1<t<t_2$).

However, in some classes of waves, which are considered below, the averaging over one variable, $t$ or $x$, appears to be sufficient.

\subsection{Waves in a homogeneous waveguide}
For waves depending only on one variable, $\Gn=x-vt$, propagating in such a waveguide, the averaging over $t$- or $x$-period, that is the averaging over $\Gn$-period, is sufficient. So, in this case, the above relation \eq{007} is valid with respect to any cross-section of the waveguide.

\subsubsection{Nonlinearly elastic beam}
\begin{figure}[h!]

\vspace{-2mm}
\hspace{30mm}\includegraphics*[width=.6\textwidth]{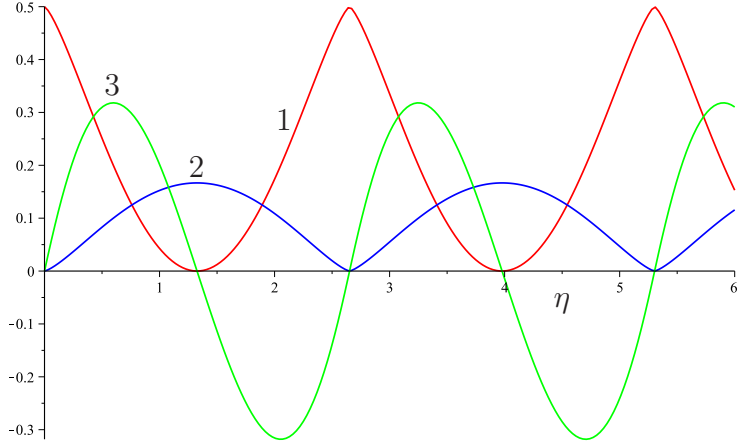}

\begin{picture}(0,0)(0,-100)
\put(157,17){2}
\put(190,35){1}
\put(125,48){3}
\put(295,-32){$\Gn$}
\end{picture}

\vspace{-4mm}
\caption[]{Periodic wave in a nonlinearly elastic beam. Kinetic energy (1), potential energy (2)  and the energy distribution function $\CL(t)$ (3) are calculated for $\nu=4, v=1, u(0)=0, \dot{u}(0)=1.$}
\label{bw4}
\end{figure}

Consider a wave in a beam where the energies are
\beq \CK= \f{\dot(u)(\Gn)^2}{2}\,,~~~\CP=\f{(u''(\Gn))^4}{4}\,,~~~\Gn=x-vt\,.\eeq{bw41}
For a periodic wave the dynamic equation
\beq v^2u''(\Gn) + \gl((u''(\Gn))^3\gr)''=0\eeq{bw42}
can be reduced to
\beq u''(\Gn) + [v^2u(\Gn)]^{1/3}=0\eeq{bw42}
The kinetic and potential energies and the energy distribution function $L(\Gn)$ calculated based on this equation are presented in \fig{bw4}.

\subsubsection{Linear and nonlinear Klein-Gordon equations}
\begin{figure}[h!]

\vspace{-2mm}
\hspace{30mm}\includegraphics*[width=.6\textwidth]{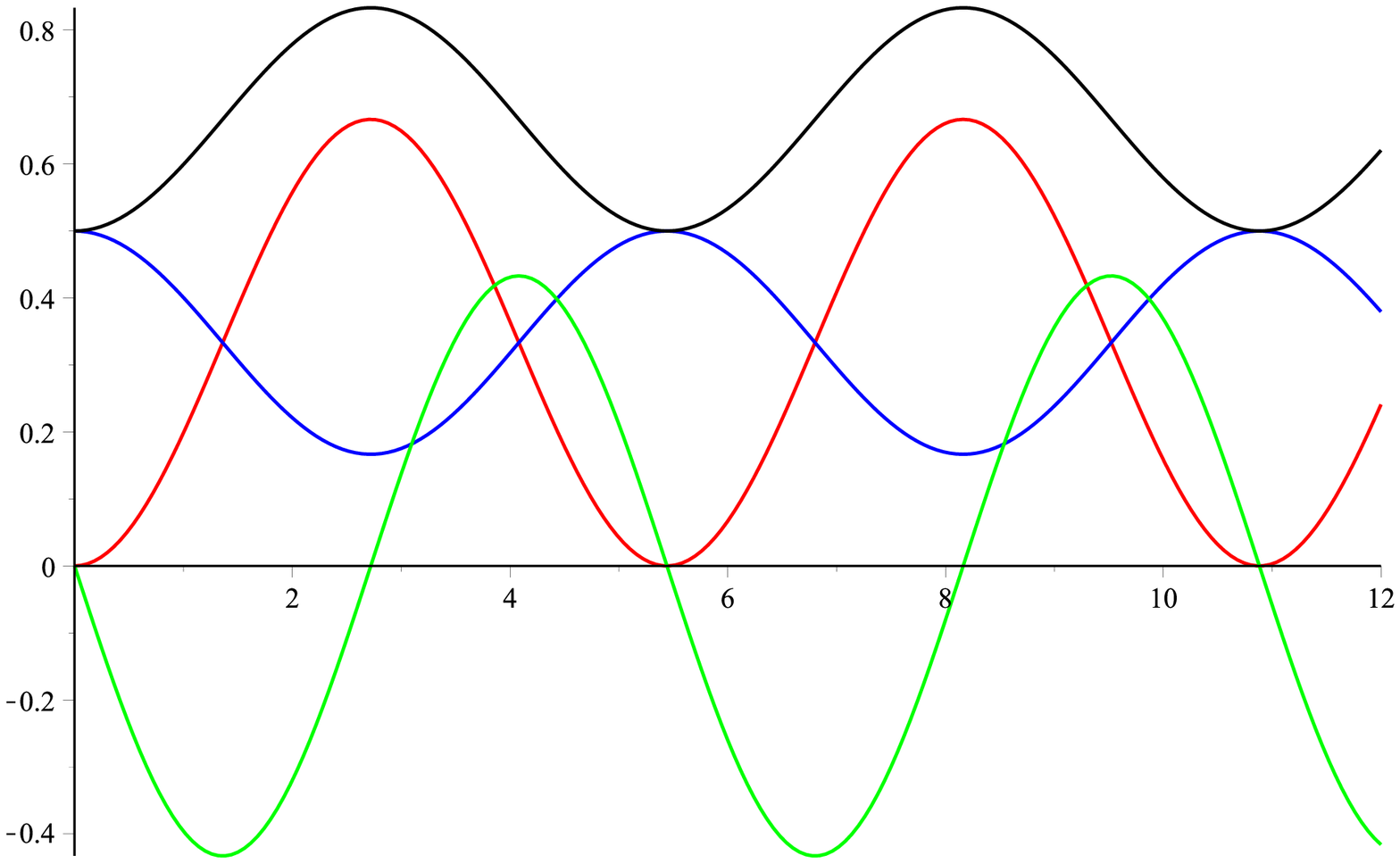}

\begin{picture}(0,0)(0,-100)
\put(119,7){1}
\put(119,35){2}
\put(119,-66){4}
\put(119,57){3}
\put(295,-41){$\Gn$}
\end{picture}

\vspace{-4mm}
\caption[]{Klein-Gordon linear equation, $\nu=2$. Kinetic energy (1), potential energy (2), total energy (3) and the energy distribution  function $\CL(t)$ (4).}
\label{w-nu2}
\end{figure}

\begin{figure}[h!]

\vspace{5mm}
\hspace{30mm}\includegraphics*[width=.6\textwidth]{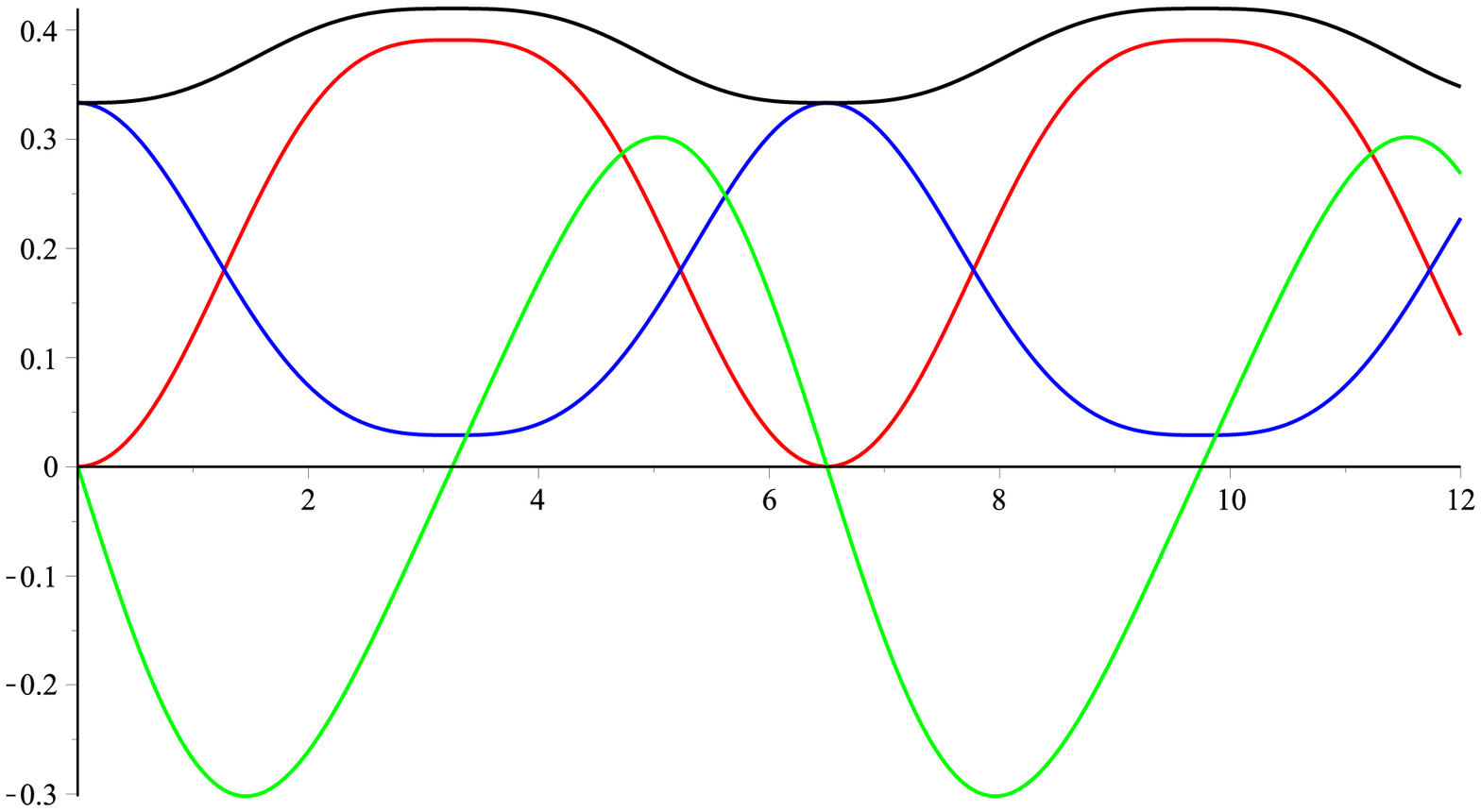}

\begin{picture}(0,0)(0,-100)
\put(119,10){1}
\put(119,38){2}
\put(119,-63){4}
\put(119,55){3}
\put(291,-32){$\Gn$}
\end{picture}

\vspace{-4mm}
\caption[]{Klein-Gordon nonlinear equation, $\nu=3$. Kinetic energy (1), potential energy (2), total energy (3) and the energy distribution function $\CL(t)$ (4).}
\label{w-nu3}
\end{figure}

\begin{figure}[h!]

\vspace{-2mm}
\hspace{30mm}\includegraphics*[width=.6\textwidth]{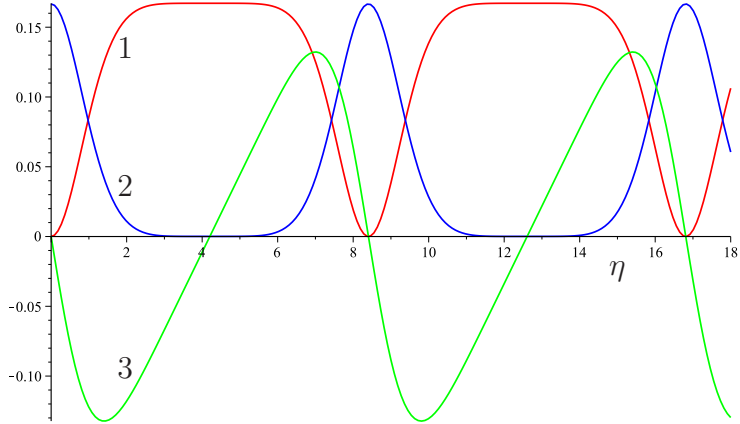}

\begin{picture}(0,0)(0,-100)
\put(130,55){1}
\put(130,3){2}
\put(130,-66){3}
\put(316,-27){$\Gn$}
\end{picture}

\vspace{-4mm}
\caption[]{Klein-Gordon nonlinear equation, $\nu=6$. Kinetic energy (1), potential energy (2) and the energy distribution  function $\CL(t)$ for 8th cell (3).}
\label{w-nu6}
\end{figure}

Consider periodic waves, $u(\Gn), \Gn =x-vt$, with the energies
\beq \CK=\f{\dot{u}^2(\Gn)}{2} = \f{v^2 (u')^2(\Gn)}{2}\,,~~~\CP=\f{|u'|^\nu(\Gn)}{\nu} + \f{|u|^\nu(\Gn)}{\nu}\,.\eeq{eow1}
The corresponding equation is
\beq \gl(v^2 - (\nu-1)|u'|^{\nu-2}(\Gn)\gr)u''(\Gn) +|u|^{\nu-2}(\Gn)u(\Gn)=0\,.\eeq{eow2}
The energies and the energy distribution function $\CL(x)$ are plotted for $\nu = 2$ (the linear wave), $3$ and $6$ in \fig{w-nu2} $-$ \fig{w-nu6}, \res. Note that, in the case of waves, the total energy varies during the period.

\subsection{Wave in periodic structures}
For a periodic structure we consider Floquet-Bloch wave, where the displacements at discrete points $x=an+x_0$, $0\le x_0<a$, depend on $x_0$ and $\Gn = an-vt$, and possibly other variables, but not on integer $n$ and time separately (here $a$ is the spatial period). This means that each of the cross-sections with the same $x_0$ describes the same periodic trajectory different only by a shift in time. We assume that (generalised) displacements of the cross-sections corresponding to a fixed $x_0$, say, $x_0=0$, defined those for the other cross-sections. 

In the considered case of an infinite waveguide, the energy of a periodic wave is also infinite. However, in the case of a finite radius of the interaction, where the $n$th cross-section is directly connected with a finite number of the others (with the same $x_0$), the displacement $\bfm{u}_n$ appears only in a finite number of the infinite series representing the Lagrangian. Accordingly, the averaging  \eq{007}  should correspond to the structure period
\beq \int_{t_1}^{t_2}\int_a \nu_mL_m\D x\D t=0\,,\eeq{wps1}
where $a$ is the spatial period of the structure. Note that the spatial integral becomes a sum in the case of a discrete structure.

As an example consider a linear mass-spring chain. The (non-dimensional) energies are
\beq \CK=\f{1}{2}\dot{u}_n^2\,,~~~\CP=\f{1}{2}(u_{n+1}-u_n)^2+\f{1}{2}(u_{n-1}-u_n)^2\,,\eeq{msdc1}
In this case, $\Gm=\nu=2$ and there is equipartition. This result also follows from the  dynamic equation
\beq \ddot{u}_n=u_{n+1}+u_{n-1}-2u_n\,,\eeq{msdc2}
which is satisfied by the wave with the corresponding dispersion relation
\beq u_n = \exp[\I(\Go t-kn)]\,,~~~\Go=\pm 2\sin\f{|k|}{2}\,.\eeq{msdc3}
From this and \eq{msdc1} we find
\beq \CK=\f{\Go^2}{2}=\CP=2\sin^2\f{k}{2}\,.\eeq{msdc4}

Wave in a nonlinear chains are considered in the next section.

\subsection{Solitary wave}
In the case of a solitary wave, the partition of its total energy is considered, and the integration limits become infinite, $t_{1,2}=\mp\infty$.

\subsubsection{Phi-Four equation}\label{P4}
Consider the so-called Phi-Four equation (see, e.g., Dodd et al (1984), Zwillinger (1997))
\beq \ddot{u}(x,t) - u''(x,t)-u(x,t)+u^3(x,t)=0\,,\eeq{ph4e1}
where dots and primes mean the derivatives with respect to $t$ and $x$, \res. This equation is satisfied by a kink
\beq u(x,t)=U(\Gn)=\pm \tanh \gl(\f{\Gn}{\sqrt{2(1-v^2)}}+\Gf_0\gr)\,,~~~\Gn=x=vt\,,\eeq{ph4e2}
where the arbitrary constants, $v\, (v^2<1)$ and $\Gf_0$, are the kink's speed and the `initial' phase, \res.
The corresponding energies are
\beq \CK=\f{\dot{u}^2(\Gn)}{2}=v^2\f{(u'(\Gn))^2}{2}\,,~~~\CP= \CP_1+\CP_2\,,\n
\CP_1=\f{(u'(\Gn))^2}{2} -\f{u^2(\Gn)}{2}\,,~~~\CP_2=\f{u^4(\Gn)}{4}\,.\eeq{ph4e3}

While the moving kink corresponds to the transition from one state to another, the particle velocity
\beq \dot{u}(\Gn)=\mp \f{v}{\sqrt{2(1-v^2)}}\gl[\cosh^2\gl(\f{\Gn}{\sqrt{2(1-v^2)}}+\Gf_0\gr)\gr]^{-1}\,,\eeq{ph4e4}
represents a true solitary wave. As a result, $u\dot{u}\to 0~ (\Gn\to\pm\infty)$ and our statement is valid with respect to the total energies of the wave
\beq \inti \CL(\Gn)\D\Gn =0\,,~~~\CL = 2(\CK - \CP_1) - 4\CP_2\,.\eeq{ph4e5}

\begin{figure}[h!]

\vspace{-2mm}
\hspace{30mm}\includegraphics*[width=.3\textwidth]{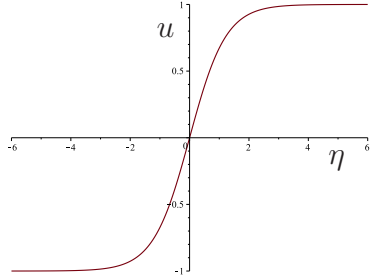}

\begin{picture}(0,0)(0,-100)
\put(144,7){$u$}
\put(209,-41){$\Gn$}
\end{picture}

\vspace{-4mm}
\caption[]{The kink.}
\label{kink}
\end{figure}

\begin{figure}[h!]

\vspace{12mm}
\hspace{20mm}\includegraphics*[width=.5\textwidth]{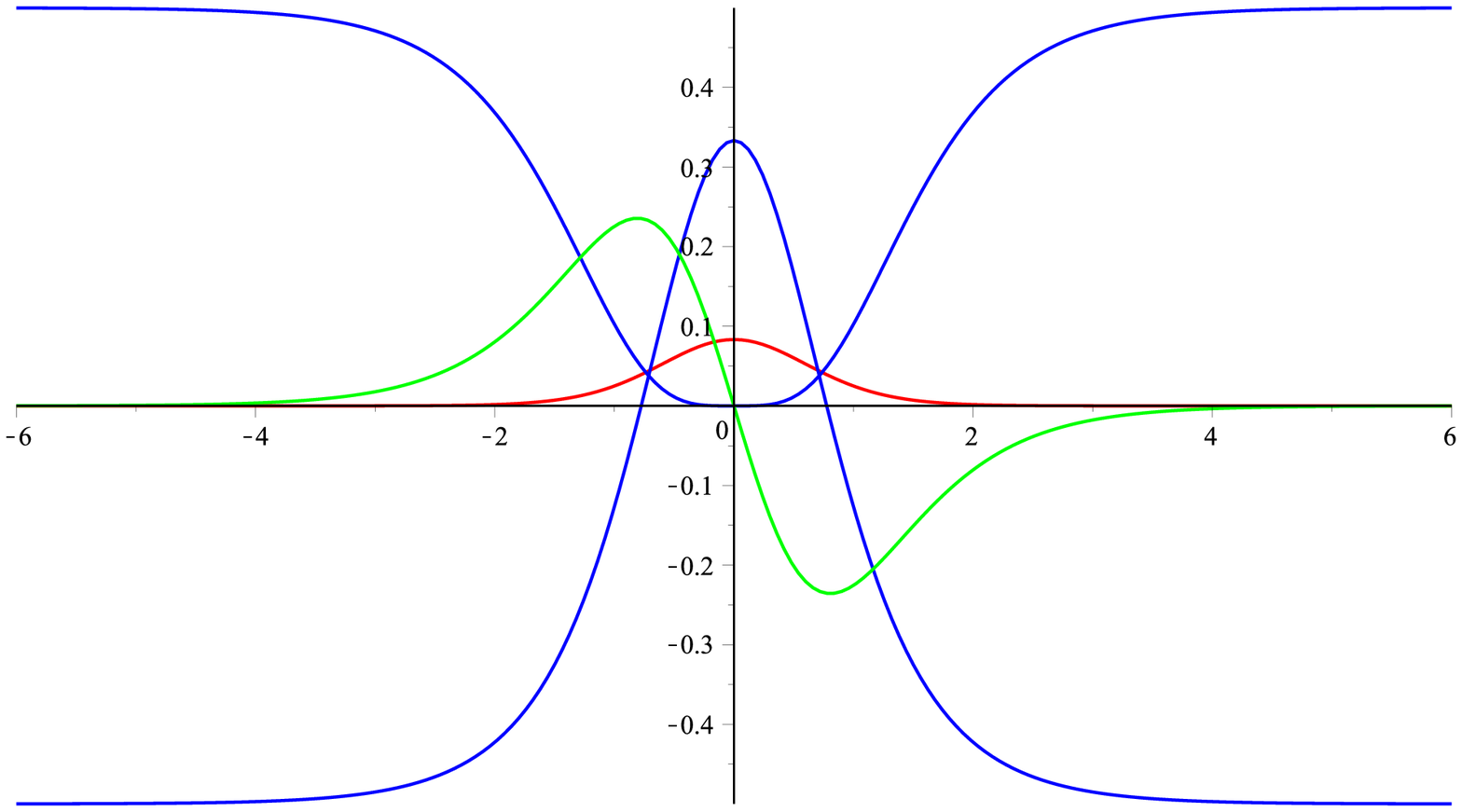}

\begin{picture}(0,0)(0,-100)
\put(178,-9){$1$}
\put(138,-63){$2$}
\put(138,27){$3$}
\put(138,-3){$4$}
\put(270,-32){$\Gn$}
\end{picture}

\vspace{-4mm}
\caption[]{Functions $\CK$ (1), $\CP_1$ (2), $2\CP_2$ (3) and the energy distribution function $\CL(x)$ (4) for the kink.}
\label{phi-fourl}
\end{figure}

\begin{figure}[h!]

\vspace{12mm}
\hspace{20mm}\includegraphics*[width=.8\textwidth]{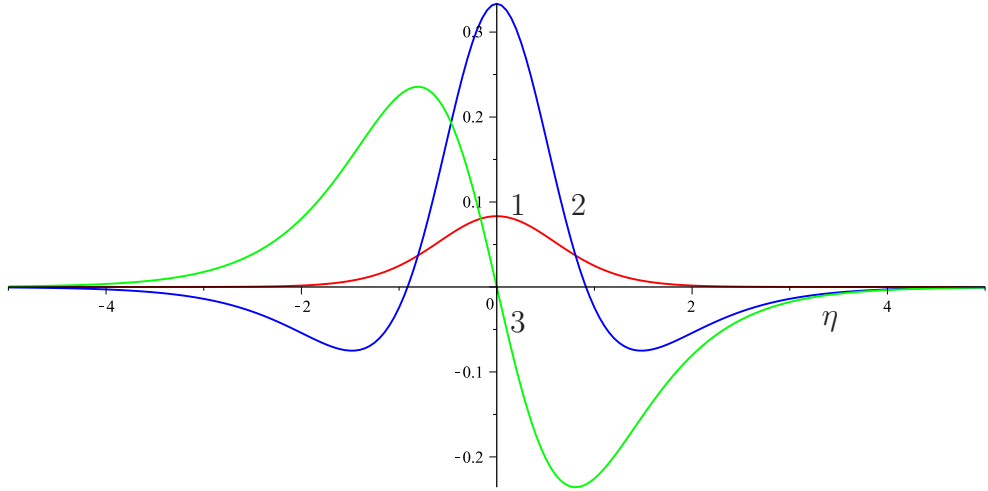}

\begin{picture}(0,0)(0,-100)
\put(250,21.5){1}
\put(273,21.5){2}
\put(250,-23){3}
\put(368,-20){$\Gn$}
\end{picture}

\vspace{-4mm}
\caption[]{The energies, $\CK$ (1), $P_1+2P_2$ (2) and the energy distribution function $\CL(x)$ (3) for the kink.}
\label{phi-fourl1}
\end{figure}

The kink, the solitary wave and the energies as functions of $\Gn$ are shown in \fig{kink} $-$ \fig{phi-fourl1}. Far away from the transition region there is a static state, where, in accordance with \eq{sccp0} and \eq{ph4e5}, $\CP_1= - u^2(\Gn)/2 = - 2\CP_2$.

\subsubsection{FPU problem}
\begin{figure}[h!]

\vspace{-0mm}
\hspace{30mm}\includegraphics*[width=.5\textwidth]{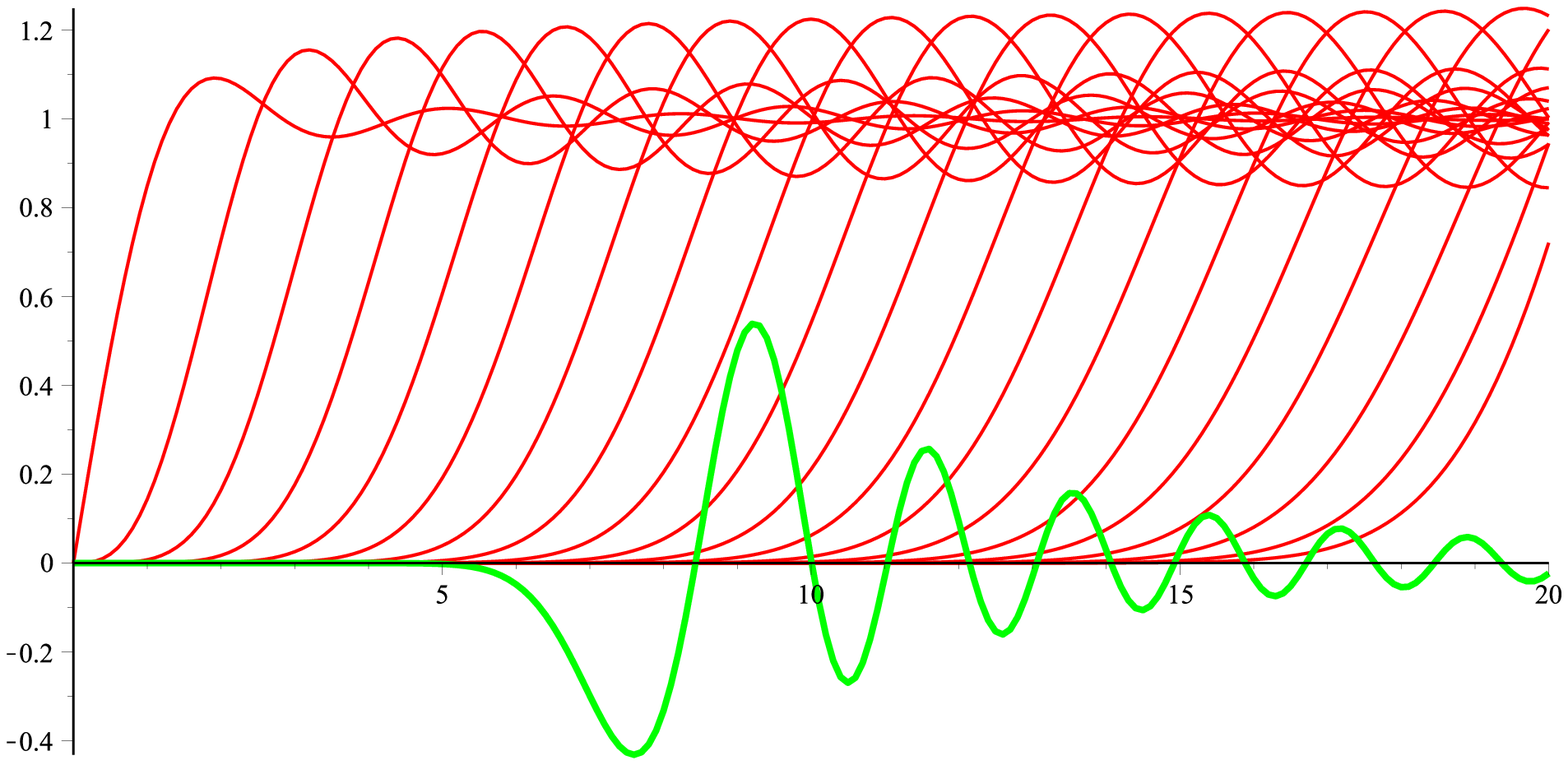}

\begin{picture}(0,0)(0,-100)
\put(129,23){1}
\put(176,-75){2}
\put(282,-64){$\Gn$}
\end{picture}

\vspace{-5mm}
\caption[]{Transient wave in the linear chain under a pulse. The displacements of the masses (1) and $20\CL(t)$ for $8$th cell (2)}
\label{FPUld}
\end{figure}

\begin{figure}[h!]

\vspace{-5mm}
\hspace{30mm}\includegraphics*[width=.6\textwidth]{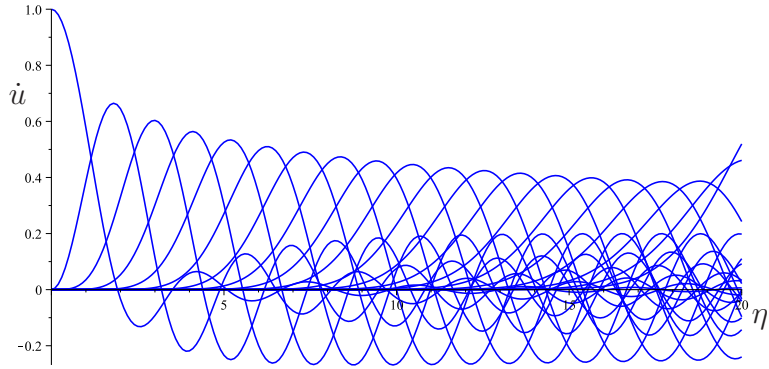}

\begin{picture}(0,0)(0,-100)
\put(85,16){$\dot{u}$}
\put(366,-67){$\Gn$}
\end{picture}

\vspace{-5mm}
\caption[]{Transient wave in the linear chain under a pulse. The decreasing wave of particle velocities.}
\label{FPUlv}
\end{figure}

Next, we consider a transient problem for a wave excited by a pulse applied to the same but `semi-infinite' linear chain. No solitary wave can exist in the latter, and we study this problem to compare the result with that for a nonlinear chain first examined by Fermi, Pasta and Ulam [8]. The displacements of the first twenty masses under unit pulse acting on the first left mass and also the corresponding speeds are presented in \fig{FPUld} and  \fig{FPUlv}, \res.

\begin{figure}[h!]

\vspace{-2mm}
\hspace{30mm}\includegraphics*[width=.5\textwidth]{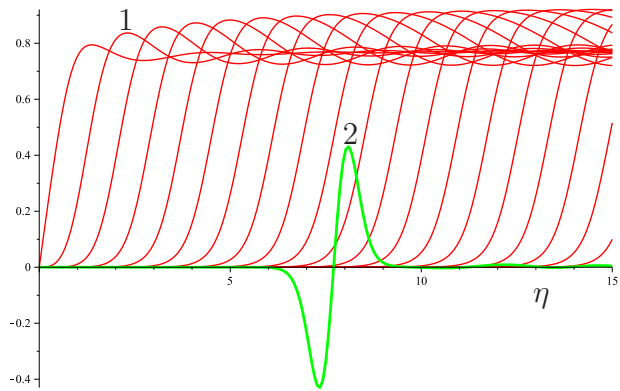}

\begin{picture}(0,0)(0,-100)
\put(129,53){1}
\put(214,10){2}
\put(286,-50){$\Gn$}
\end{picture}

\vspace{-5mm}
\caption[]{Transient wave in the nonlinear chain under a pulse (FPU problem). The displacements of the masses (1) and $8\CL(t)$ for 10th cell (2).}
\label{FPUnd}
\end{figure}

\begin{figure}[h!]

\vspace{-2mm}
\hspace{30mm}\includegraphics*[width=.5\textwidth]{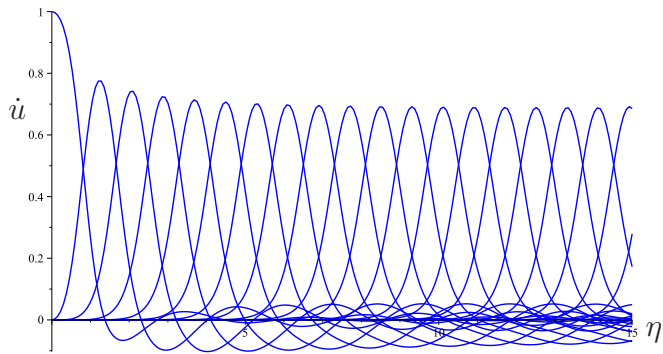}

\begin{picture}(0,0)(0,-100)
\put(80,5){$\dot{u}$}
\put(321,-77){$\Gn$}
\end{picture}

\vspace{-4mm}
\caption[]{Transient wave in the nonlinear chain under a pulse (FPU problem). The establishing solitary wave of particle velocities.}
\label{FPUns}
\end{figure}

The wave itself, especially the particle velocity wave, and the energy distribution function, $\CL(t)$, can be compared with those  in the FPU problem [8] for the same but nonlinear chain, \fig{FPUnd} and \fig{FPUns}, where a stable solitary wave arises (compare \fig{FPUlv} with \fig{FPUns}), and with `pure nonlinear' one corresponding to the potential energy without linear terms.
In a version of the FPU problem, the energies are taken as
\beq \CK=\f{1}{2}\dot{u}_n^2\,,~~~\CP=\f{1}{2}(u_{n+1}-u_n)^2+\f{1}{2}(u_{n-1}-u_n)^2+(u_{n+1}-u_n)^4+(u_{n-1}-u_n)^4\,,\eeq{msdc1n}
Here $\Gm=\nu_1=2, \nu_2=4$ and $\inti (\CK - \CP_1 -2\CP_2)\D t =0$.

\begin{figure}[h!]

\vspace{-2mm}
\hspace{30mm}\includegraphics*[width=.6\textwidth]{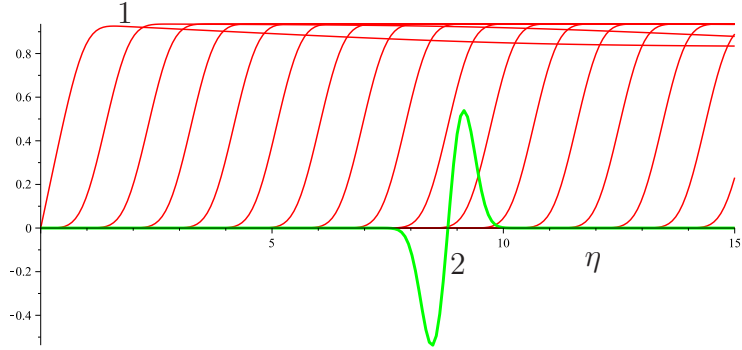}

\begin{picture}(0,0)(0,-100)
\put(129,40){1}
\put(255,-55){2}
\put(306,-52){$\Gn$}
\end{picture}

\vspace{-4mm}
\caption[]{Solitary wave in the nonlinear chain with no quadratic terms in the expression of the potential energy \eq{msdc1n}. The displacements of the masses (1) and $4\CL(t)$ for 10th cell (2).}
\label{FPUnnd}
\end{figure}

\begin{figure}[h!]

\vspace{-2mm}
\hspace{30mm}\includegraphics*[width=.5\textwidth]{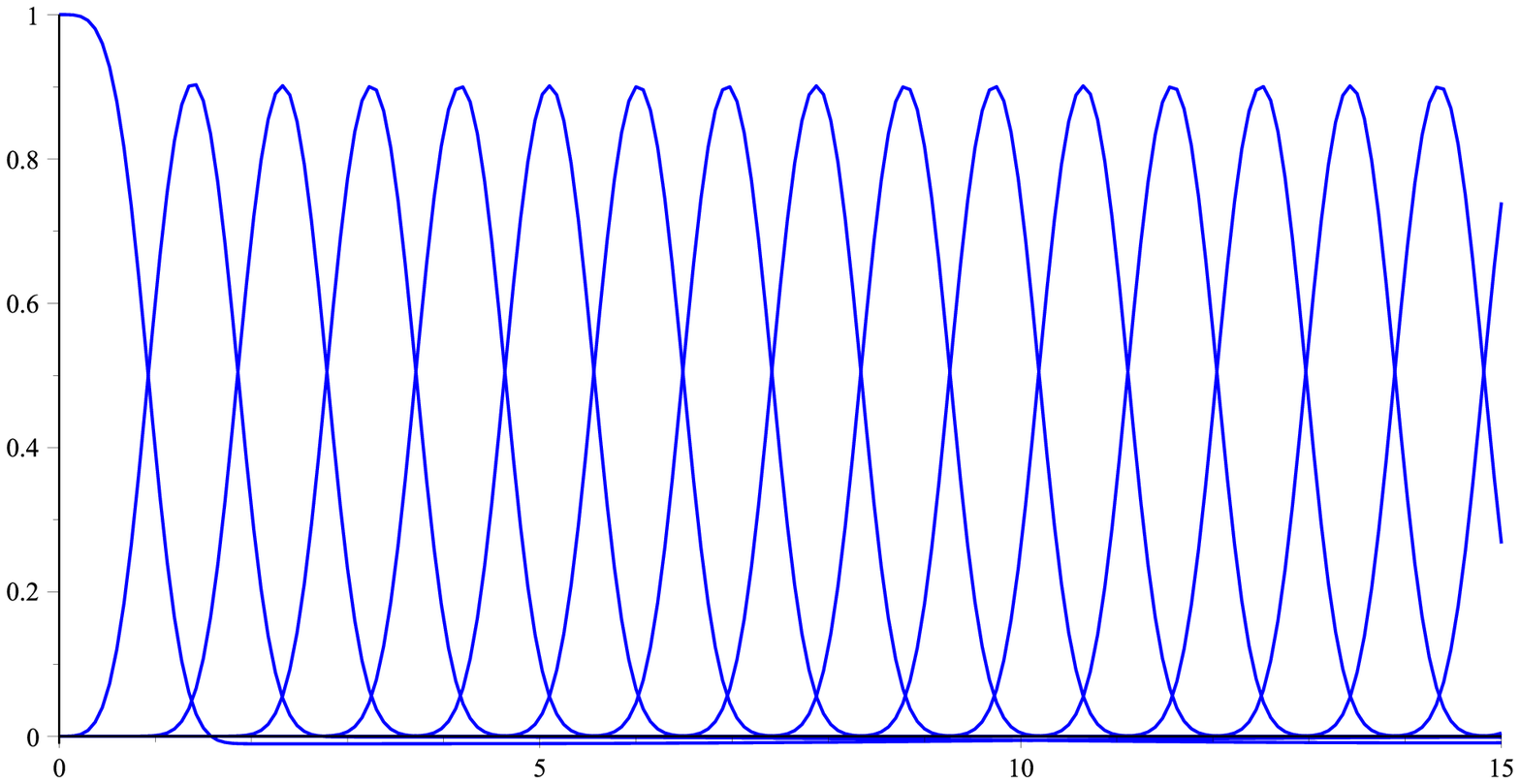}

\begin{picture}(0,0)(0,-100)
\put(80,-5){$\dot{u}$}
\put(291,-87){$\Gn$}
\end{picture}

\vspace{-4mm}
\caption[]{Solitary wave in the nonlinear chain with no quadratic terms in the expression of the potential energy \eq{msdc1n}. The particle velocities.}
\label{FPUnns}
\end{figure}

A fast-established strongly localized solitary wave appears in the nonlinear chain with quadratic terms in the expression of the potential energy \eq{msdc1n} removed. In this case, $\CK=2\CP$. The calculation results are presented in \fig{FPUnnd} and \fig{FPUnns}

\section{Conclusions}
The energy partition obeys the relation derived for the case where the Lagrangian is represented as a sum of homogeneous functions of the displacements and their derivatives (explicit dependencies on time and spacial coordinates are not excluded). If the kinetic and potential energies are entirely homogeneous the homogeneity orders define the partition  uniquely.

The equipartition corresponds to the case where the kinetic and potential energies are of the same homogeneity order. In particular, this is true for homogeneous linear problems where both energies are of the second order. At the same time, the linearity is neither necessary nor sufficient condition for the equipartition.

The energy partition corresponds to the energies averaged over a region satisfying the condition \eq{005}. Thus, the integration can correspond not only to the period, if exists, but to end points where $\bfm{u}=0$ and, in the case \eq{012},  where $\dot{\bfm{u}}=0$. This is especially important for non-periodic oscillations and waves, which can exist in transient problems and under time-varying parameters as discussed above.  For a solitary wave the total energies are considered, and the averaging integral is taken over the whole axis (or a half-axis).

The energy partition relation is valid as far as the related general conditions are satisfied. So, it is applicable to any specific problem of the corresponding class, and there is no need in the examination of any specific problem in more detail. Nevertheless, we have presented various examples with the aim to show how the partition varies during the time-interval of averaging, and also to demonstrate validity of the relation and the conditions relating to the averaging, These examples show that the relation does valid for various problems, linear and nonlinear, steady-state and transient, conservative and non-conservative, homogeneous and forced oscillations, periodic and solitary waves.

\vspace{15mm}
\noindent The author {\bf acknowledges} the support provided by the FP7 Marie Curie grant  No. 284544-PARM2.

\vspace{10mm}
\vskip 18pt
\begin{center}
{\bf  References}
\end{center}
\vskip 3pt

      1. Antenucci, J.P., and Imburger, J.,2001. Energetics of long internal gravity waves in large lakes. Limnol. Oceanogr., 46(7), 1760–1773.

      2. Falnes, J., 2007.  A review of wave-energy extraction. Marine Structures 20, 185–201.

      3. Whitham, G.B., 1974. Linear and Nonlinear Waves. John Wiley \& Sons, NY.

      4. Thiago Messias Cardozo and Marco Antonio Chaer Nascimento, 2009. Energy partitioning for generalized product functions: The interference contribution to the energy of generalized valence bond and spin coupled wave functions. The Journal of Chemical Physics 130, 104102-1-8.

      5. Korycansky, D.G., 2013. Energy conservation and partition in CTH impact simulations. Lunar and Planetary Science Conference, 1370.pdf.

      6. Dodd,R.K., Eilbeck, J.C., Gibbon, J.D., Morris, H.C., 1984. Solitons and Nonlinear Wave Equations. Academic Press, NY.

      7. Zwillinger, D., 1997. Handbook of Differential Equations, 3rd ed. Boston, MA: Academic Press.

      8. Fermi E., Pasta J. and Ulam S., 1955. Studies of nonlinear problems. I. Los Alamos report LA-1940, published later in Collected Papers of Enrico Fermi, E. Segré (Ed.), University of Chicago Press (1965).

\end{document}